\documentclass[prl,reprint,superscriptaddress,showpacs,
amsmath,amssymb,floatfix]{revtex4-1}
\pdfoutput=1

\usepackage{graphicx} % Include figure files
\usepackage{dcolumn} % Align table columns on decimal point
\def\bG{{\bf G}}

\def\bk{{\bf k}}

\def\bq{{\bf q}}

\def\bG{{\bf G}}
\def\bQ{{\bf Q}}

\def\b0{{\bf 0}}

\def\Im{{\rm Im}}

\def\lra{\leftrightarrow}

\def\eps{\epsilon}

\def\om{\omega}

\def\Sg{\Sigma}

\def\tk{\tilde k}
\def\tom{\tilde\om}

\def\sgn{{\rm sgn}}

\begin{document}

\title{Non-Fermi liquid behavior at the onset of incommensurate \\
 $2k_F$ charge or spin density wave order in two dimensions}

\author{Tobias Holder}
\affiliation{Max Planck Institute for Solid State Research,
 D-70569 Stuttgart, Germany}
\author{Walter Metzner}
\affiliation{Max Planck Institute for Solid State Research,
 D-70569 Stuttgart, Germany}

\date{\today}

\begin{abstract}
We analyze the influence of quantum critical fluctuations on 
single-particle excitations at the onset of incommensurate $2k_F$
charge or spin density wave order in two-dimensional metals.
The case of a single pair of hot spots at high symmetry positions
on the Fermi surface needs to be distinguished from the case of
two hot spot pairs.
We compute the fluctuation propagator and the electronic self-energy
perturbatively in leading order.
The energy dependence of the single-particle decay rate at the 
hot spots obeys non-Fermi liquid power laws, with an exponent 2/3 
in the case of a single hot spot pair, and exponent one for two hot 
spot pairs. The prefactors of the linear behavior obtained in the 
latter case are not particle-hole symmetric.
\end{abstract}

\maketitle

%%% Intro %%%%%%%%%%%%%%%%%%%%%%%%%%%%%%%%%%%%%%%%%%%%%%%%%%%%%%%%

Charge and spin correlations in metals exhibit a well-known
singularity at wave vectors that connect points on the Fermi 
surface with antiparallel Fermi velocities.
The singularity is caused by an enhanced phase space for
low-energy particle-hole excitations near such wave vectors.
It leads, among other effects, to the Kohn anomaly \cite{kohn59} 
in phonon spectra and to the long-ranged RKKY interaction between 
magnetic impurities in metals \cite{rkky}.
For isotropic Fermi surfaces the singularity is located at 
wave vectors with modulus $2k_F$, where $k_F$ is the radius 
of the Fermi surface.
In inversion symmetric crystalline solids, singular wave vectors
are given by the condition
\begin{equation} \label{2kf}
 \xi_{(\bQ+\bG)/2} = 0 \; ,
\end{equation}
where $\xi_{\bk} = \eps_{\bk} - \mu$ is the single-particle
excitation energy, and $\bG$ is a reciprocal lattice vector.
Eq.~(\ref{2kf}) is the lattice generalization of the condition
$|\bQ| = 2k_F$ for isotropic systems. Hence, we generally refer
to wave vectors satisfying that condition as $2k_F$ vectors.

$2k_F$ singularities are more pronounced in systems with 
reduced dimensionality. Charge and spin correlations in 
low dimensional systems are often peaked at $2k_F$ vectors,
such that they are privileged wave vectors for charge and
spin density wave instabilities. $2k_F$ instabilities are
ubiquitous in (quasi) one-dimensional electron systems
\cite{giamarchi04}.
Here we focus on {\em two-dimensional}\/ systems, where 
$2k_F$ instabilities also play an important role.
In particular, the ground state of the two-dimensional Hubbard 
model exhibits a spin density wave instability at a $2k_F$ 
vector, at least at weak coupling \cite{schulz90,igoshev10}.
Furthermore, spatially modulated nematic order, that is,
$d$-wave bond charge order, occurs preferably at $2k_F$ vectors
\cite{holder12}. 
For Fermi surfaces crossing the antiferromagnetic zone 
boundary, the highest peaks in the $d$-wave charge response 
are at $2k_F$ vectors connecting magnetic hot spots, such 
that the bond order instability can be triggered by 
antiferromagnetic interactions \cite{metlitski10_af,sachdev13}.

In this paper we analyze consequences of quantum criticality
at the onset of {\em incommensurate}\/ $2k_F$ charge or spin 
density wave order in the ground state, in cases where the 
phase transition (e.g., as a function of electron density) is 
continuous.
The momentum and energy dependences of the $2k_F$ fluctuation 
propagator differs strongly from that for generic incommensurate
wave vectors \cite{castellani95}, 
and also from the one for commensurate $(\pi,\pi)$ charge or spin 
density wave instabilities \cite{hertz76,millis93}, for which 
quantum critical properties have been extensively studied 
\cite{abanov03,metlitski10_af}.
The quantum critical behavior at $2k_F$ density wave 
transitions in two dimensional metals was addressed many 
years ago by Altshuler et al.\ \cite{altshuler95}, who
computed several properties for the case that the $2k_F$
vector is half a reciprocal lattice vector. The special case
where $(\pi,\pi)$ is a $2k_F$ vector was recently revisited,
and qualitative modifications due to additional umklapp
processes were revealed \cite{bergeron12}. 
For incommensurate $2k_F$ vectors, Altshuler et al.\ found 
strong infrared divergencies and concluded that fluctuations 
destroy the quantum critical point (QCP), such that the phase 
transition is ultimately discontinuous.

In the following we will analyze the influence of incommensurate
$2k_F$ quantum critical fluctuations on single-particle 
excitations by computing the electronic self-energy at $2k_F$ hot
spots \cite{fn1} on the Fermi surface to first order (one loop) 
in the fluctuation propagator. 
We will show that one needs to distinguish the case where the
$2k_F$ vector connects only one pair of hot spots at high
symmetry points from cases where it connects two hot spot pairs. 
Only the former case was considered in Ref.~\cite{altshuler95}. 
In both cases, the quasi-particle decay rate obeys non-Fermi
liquid power laws as a function of energy, but with distinct
exponents, $2/3$ and one, respectively.
These power laws may be observed in a certain energy window 
even in case that the fluctuations are ultimately cut off by 
a first order transition, or by a secondary instability in 
close vicinity of the QCP.

%%% Bubble and effective interaction %%%%%%%%%%%%%%%%%%%%%%%%%

To compute the electronic self-energy, we first need to derive
the momentum and energy dependence of the effective interaction 
(fluctuation propagator) at the QCP. 
At leading order, the latter is given by the RPA expression 
$D(\bq,\om) = g \left[ 1 - g \Pi_0(\bq,\om) \right]^{-1}$,
where $g$ is the coupling parametrizing the bare interaction 
in the instability channel, and $\Pi_0$ is the bare polarization
function of the system. The RPA effective interaction and the
one-loop self-energy are not affected qualitatively by details 
such as the spin structure and form factors ($s$-wave, $d$-wave, 
etc.) in $\Pi_0(\bq,\om)$.
Finite renormalizations from non-critical fluctuations could 
be incorporated by using a renormalized coupling and a reduced 
quasi-particle weight.

At the onset of density wave order, $g \Pi_0(\bQ,0)$ is equal
to one such that the effective interaction diverges.
The momentum and energy dependence of $D(\bq,\om)$ at that
point is obtained by expanding $\Pi_0(\bq,\om)$ for $\bq$ near
$\bQ$ and small energies. Here the $2k_F$-singularity comes
into play. 
The shape of that singularity can be deduced from the expanded
analytic result \cite{stern67} for $\Pi_0(\bq,\om)$ for fermions 
in the continuum with a parabolic dispersion 
$\eps_{\bk} = \frac{\bk^2}{2m}$,
\begin{eqnarray} \label{stern}
 \Pi_0(\bq,\om) &=& - \frac{m}{2\pi} + \frac{\sqrt{m}}{4\pi v_F}
 \nonumber \\ 
 &\times& \left( 
 \sqrt{e_{\bq} + \om + i0^+} +  \sqrt{e_{\bq} - \om - i0^+}
 \right) , \hskip 5mm
\end{eqnarray}
where $e_{\bq} = v_F(|\bq| - 2k_F)$, and $v_F$ is the Fermi 
velocity.
The complex roots are defined with a branch cut on the negative
real axis. The infinitesimal imaginary parts under the roots
specify that the real frequency (that is, energy) axis is 
approached from above.
For electrons in a crystal, $\Pi_0(\bq,\om)$ is singular on the 
$2k_F$-lines given by $\xi_{(\bq+\bG)/2} = 0$.
The momentum dependence is regular along these lines, but
singular in perpendicular direction.
To parametrize momenta near the instability vector $\bQ$, we
use normal and tangential coordinates $q_r$ and $q_t$, 
respectively, as described in Fig.~1. 
\begin{figure}[htb]
\begin{center}
\includegraphics[width=3.5cm]{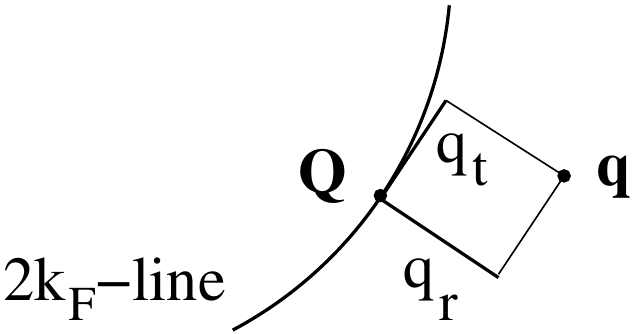}
\caption{Local coordinates $q_r$ and $q_t$ for momenta near 
 $\bQ$ in the $\bq$-plane.}
\end{center}
\end{figure}
The coordinate $q_r$ is defined positive on the ``outer'' side
of the $2k_F$-line and negative on the ``inner'' side.
The generalization of Eq.~(\ref{stern}) for electrons in a 
crystal can then be written as
\begin{eqnarray} \label{pi0}
 \Pi_0(\bq,\om) &=& \Pi_0(\bQ,0) + 
 a \Big( \sqrt{e_{\bq} + \om + i0^+} 
\nonumber \\
 &+&  \sqrt{e_{\bq} - \om - i0^+}
 \Big) - b \, e_{\bq} - c \, q_t 
\end{eqnarray}
for $\bq$ near $\bQ$ and small $\om$, where $a,b,c$ are 
constants and $e_{\bq} = v_F q_r + \frac{q_t^2}{4m}$.
Here $v_F$ is the Fermi velocity at the Fermi points $\bk_F$
and $-\bk_F$ connected by $\bQ$, and the effective mass $m$ 
parametrizes the curvature of the Fermi surface at those
points ($m v_F$ is the radius of curvature).
Note that $e_{\bq}/v_F$ is the oriented distance of $\bq$
from the $2k_F$-line in the $\bq$-plane, defined positive
outside and negative inside.
The prefactor of the singularity $a$ is fully determined by
the Fermi velocity and curvature at $\pm \bk_F$ as
\begin{equation}
 a = N \frac{\sqrt{m}}{4\pi v_F} \; ,
\end{equation}
where $N$ is the number of internal degrees of freedom such
as spin. If $\Pi_0$ contains form factors, there may be a
corresponding additional factor.
The other constants $b$ and $c$ receive contributions from
everywhere; $b$ vanishes for a quadratic dispersion, but is
otherwise typically positive; $c$ vanishes for any isotropic
dispersion, and also at symmetry points on the lattice, that
is, for an axial or diagonal $\bk_F$.
Eq.~(\ref{pi0}) describes the singularity at $2k_F$-vectors
$\bQ$ connecting a single pair of hot spots on the Fermi
surface. 
The more complicated but important case of $2k_F$-vectors 
connecting two such pairs will be discussed below.
Inserting Eq.~(\ref{pi0}) into the RPA expression for the
effective interaction one obtains, at the QCP,
\begin{eqnarray} \label{D}
 D(\bq,\om) &=& 
 - \Big[ a \Big( \sqrt{e_{\bq} + \om + i0^+} + 
 \sqrt{e_{\bq} - \om - i0^+} \Big) 
 \nonumber \\
 &-& b e_{\bq} - c q_t \Big]^{-1} \, .
\end{eqnarray}
Note that the coupling constant $g$ has canceled out.
The effective interaction diverges for $\bq \to \bQ$, $\om \to 0$.
The momentum and energy dependence of the singularity differs
strongly from the one for density wave instabilities at generic
(not $2k_F$) wave vectors \cite{castellani95,hertz76,millis93}.

%%% Self-energy %%%%%%%%%%%%%%%%%%%%%%%%%%%%%%%%%%%%%%%%%%%%%%%%%%

The imaginary part of the one-loop self-energy for electrons
coupled by a fluctuation propagator $D(\bq,\om)$ can generally
be written as \cite{dellanna06}
\begin{eqnarray} \label{Sg_gen}
 \Im\Sg(\bk,\om) &=& M \int \frac{d^dk}{(2\pi)^d}
 \left[ b(\xi_{\bk'}-\om) + f(\xi_{\bk'}) \right]
 \nonumber \\
 &\times& \Im D(\bk'-\bk,\xi_{\bk'}-\om) \; ,
\end{eqnarray}
where $b$ and $f$ are the Bose and Fermi functions, respectively.
The multiplicity factor $M$ is one for charge fluctuations, and 
three for spin fluctuations. At zero temperature, 
\begin{equation}
 b(\xi_{\bk'}-\om) + f(\xi_{\bk'}) = \left\{
 \begin{array}{rcl}
 - 1 & \mbox{for} & 0 < \xi_{\bk'} < \om  \; , \\ 
   1 & \mbox{for} & \om < \xi_{\bk'} < 0  \; ,
 \end{array} \right.
\end{equation}
and otherwise 0.
We now compute the low energy behavior of $\Im\Sg(\bk_F,\om)$ at
hot spots $\bk_F$ on the Fermi surface.

\begin{figure}[htb]
\begin{center}
\includegraphics[width=3.5cm]{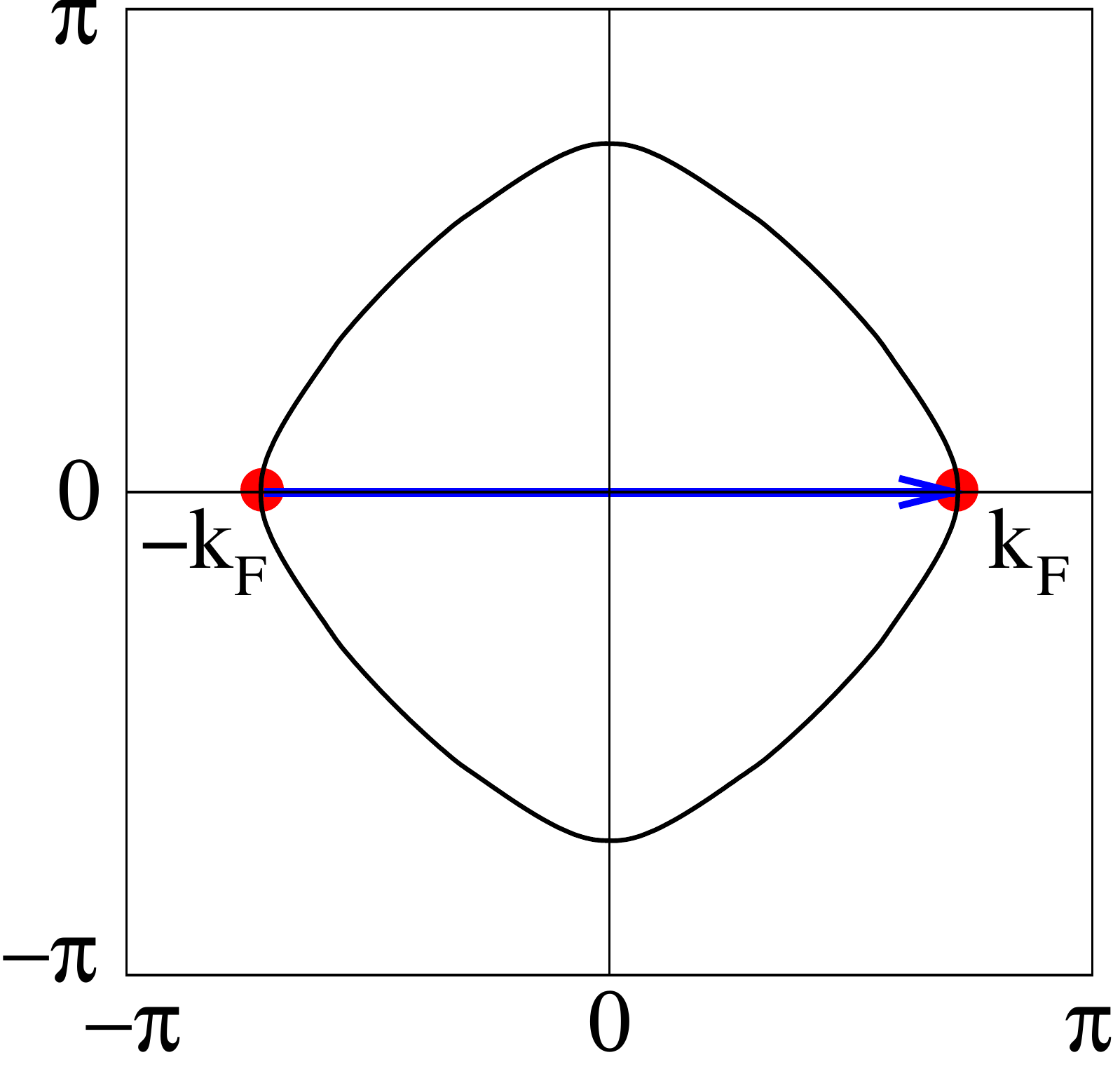} \hskip 1cm
\includegraphics[width=3.5cm]{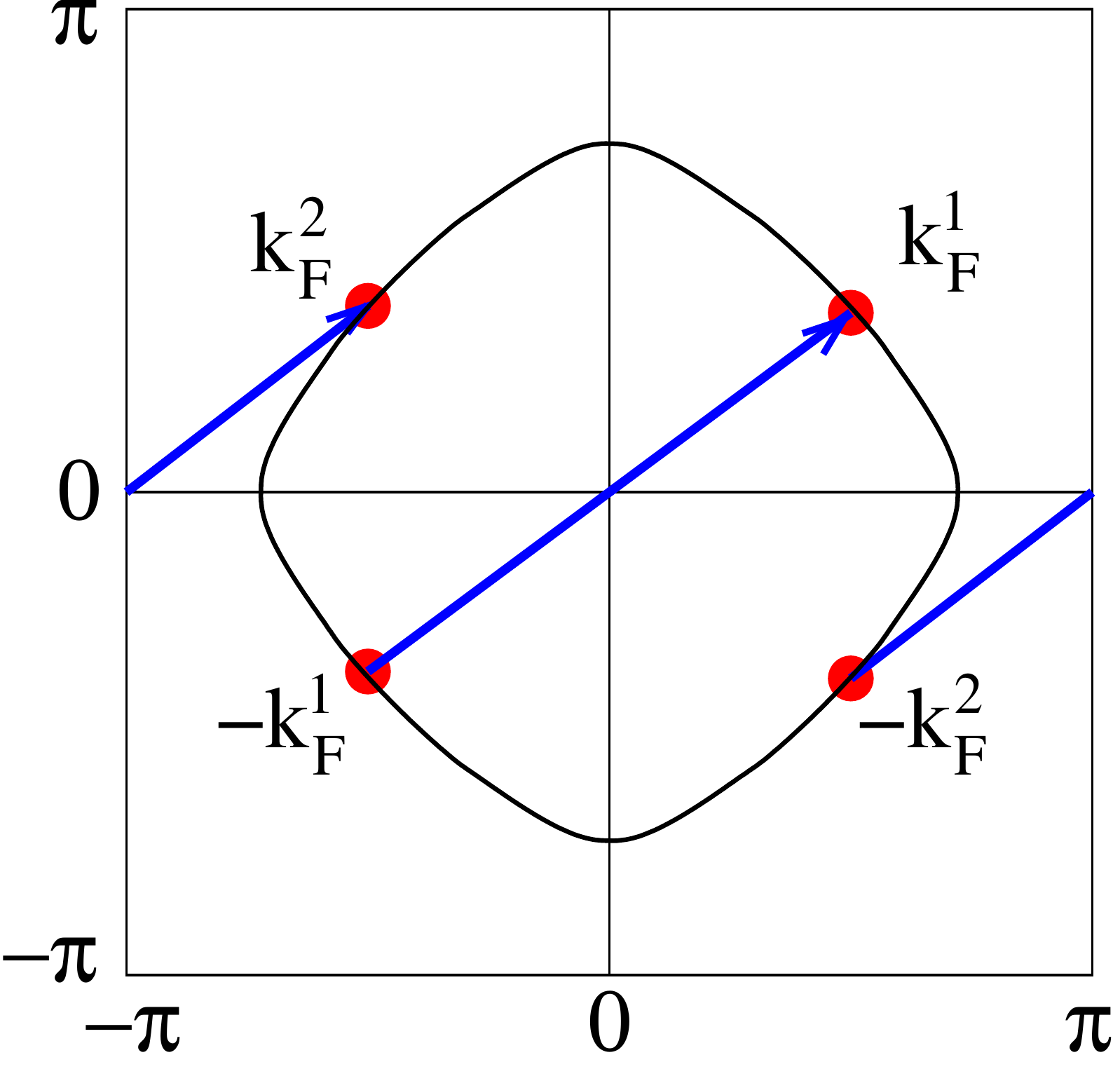}
\caption{(a) Axial $2k_F$ wave vector of the form $(Q,0)$ 
 and (b) $2k_F$ wave vector of the form $(\pi,Q)$, and the 
 corresponding hot spots on the Fermi surface.}
\end{center}
\end{figure}
$2k_F$ instabilities with a single pair of hot spots $\pm\bk_F$ occur
naturally at high symmetry points, that is, with $\bk_F$ and $\bQ$ in 
axial (see Fig.~2a) or diagonal direction.
In these cases, the fluctuation propagator at the QCP is given by 
Eq.~(\ref{D}) with $c=0$. 
For $\bk = \bk_F$ the dominant contributions to the integral in
Eq.~(\ref{Sg_gen}) come from momenta $\bk'$ near $-\bk_F$.
We assume that the Fermi surface is convex at $\pm\bk_F$ in the 
following steps, but the final result is equally valid for a concave 
Fermi surface.
Introducing normal and tangential coordinates for $\bk'$ near $-\bk_F$,
one can expand $\xi_{\bk'} = v_F k'_r + \frac{{k'_t}^2}{2m}$ and
$e_{\bk' - \bk_F} = v_F k'_r + \frac{{k'_t}^2}{4m}$.
Substituting the integration variables $k'_r$ and $k'_t$ by the new
variables $k' = - e_{\bk' - \bk_F}/v_F$ and $\om' = \xi_{\bk'}$, 
with the Jacobian $\sqrt{m}/\sqrt{\om' + v_F k'}$, one obtains
\begin{eqnarray} \label{Sg1}
 \Im\Sg(\bk_F,\om) &=& \frac{2 M v_F}{\pi N} \int_0^{\om} d\om'
 \int_{-\om'/v_F}^{\infty} \frac{dk'}{\sqrt{\om' + v_F k'}}
 \nonumber \\
 &\times& \Im \Big[
  \sqrt{\om' + i0^+ - \om - v_F k'}
 \nonumber \\ 
 &+& \sqrt{\om - \om' - i0^+ - v_F k'} + \bar b v_F k' \Big]^{-1}
\end{eqnarray}
for $\om > 0$, and a similar expression for $\om < 0$. 
The constant $\bar b$ is given by
\begin{equation}
 \bar b = b/a = \frac{4\pi v_F}{N \sqrt{m}} \, b \; .
\end{equation}
For $\om \to 0$, the integral in Eq.~(\ref{Sg1}) is dominated by 
large momenta, $k' \gg (\om - \om')/v_F$, such that we can approximate 
$\sqrt{\om' + i0^+ - \om - v_F k'} + \sqrt{\om - \om' - i0^+ - v_F k'}$
by $i(\om - \om')/\sqrt{v_F k'}$.
Substituting $k' = \om \tk/v_F$, $\om' = \om \tom$, we obtain
\begin{equation}
 \Im\Sg(\bk_F,\om) = \frac{2M \om}{\pi N} 
 \int_0^1 \! d\tom \int_{-\tom}^{\infty} \! \Im 
 \frac{d\tk}{i (1 - \tom) + \bar b \sqrt{\om} \tk^{3/2}} \, .
\end{equation}
The integral can now be computed analytically, yielding
\begin{equation}
 \Im\Sg(\bk_F,\om) =
 - \frac{2M}{\sqrt{3} N} (|\om|/\bar b)^{2/3} \; ,
\end{equation}
which is valid also for $\om < 0$.
The self-energy at the hot spots thus exhibits a non-Fermi
liquid power law behavior with an exponent $2/3$.
The leading contributions come from energies $\om' \sim \om$, 
normal momenta $k'_r \sim |\om|^{2/3}$, and tangential momenta
$k'_t \sim |\om|^{1/3}$. Remarkably, the same scaling behavior 
holds for nematic and U(1)-gauge quantum criticality 
\cite{nayak94,oganesyan01,metzner03,metlitski10_nem}, although the 
momentum and energy dependence of the fluctuation propagator
is completely different.

%%% Two pairs of hot spots %%%%%%%%%%%%%%%%%%%%%%%%%%%%%%%%%%%%%

We now turn to the second important case, where $\bQ$ connects two
pairs of hot spots. This happens for spin density instabilities
at wave vectors of the form $\bQ = (\pi,Q)$ or $(Q,\pi)$, as found for
example in Hartree-Fock calculations for the two-dimensional
Hubbard model \cite{schulz90,igoshev10}. The pairs of hot spots 
are located at $\pm \big( \frac{\pi}{2},\frac{Q}{2} \big)$ and
$\pm \big( \frac{\pi}{2},-\frac{Q}{2} \big)$ in this case (Fig.~2b).
Another example is provided by $d$-wave bond charge order at wave 
vectors of the form $\bQ = (Q,Q)$ \cite{metlitski10_af,holder12}.
The very fact that $\bQ$ connects two pairs of hot spots leads
to peaks in the polarization function, such that density wave
instabilities are favored at such special wave vectors. 
We now compute the singularity of the fluctuation propagator at the 
QCP and the energy dependence of the self-energy at hot spots for 
incommensurate density wave instabilities with two hot spot pairs. 
We present the calculation for the specific case $\bQ = (\pi,Q)$ 
shown in Fig.~2b, while the result holds for the other cases, too.

The wave vector $\bQ$ is a crossing point of two $2k_F$-lines.
Hence, the $2k_F$ singularities on these lines have to be added,
leading to
\begin{equation} \label{pi0_two}
 \Pi_0(\bq,\om) = \Pi_0(\bQ,0) + \!\! \sum_{n=1,2} \! a \left( 
 \sqrt{e_{\bq}^n + \om} + \! \sqrt{e_{\bq}^n - \om}
 \right) - b \, e_{\bq}^n \, ,
\end{equation}
where $e_{\bq}^n/v_F$ is the oriented distance from the $n$-th 
$2k_F$-line. We have suppressed the infinitesimal imaginary parts 
$i0^+$ to shorten the formula.
We have also discarded the tangential momentum dependence (that 
is, $c=0$) to simplify the expressions. It will become clear that
this term affects only prefactors and can easily be reinstalled
in the end.
The fluctuation propagator at the QCP is then given by
\begin{equation} \label{D_two}
 D(\bq,\om) = 
 - \Big[ \sum_{n=1,2} a \left( 
 \sqrt{e_{\bq}^n + \om} + \sqrt{e_{\bq}^n - \om} \right) 
 - b e_{\bq}^n \Big]^{-1} .
\end{equation}

We now evaluate the one-loop self-energy at one of the hot spots,
say $\bk_F^1$. It doesn't matter which hot spot we choose, 
because they are related by lattice symmetries. 
The integral in Eq.~(\ref{Sg_gen}) is dominated by 
momenta $\bk'$ near $-\bk_F^1$. Representing $\bk'$ by normal and
tangential coordinates $k'_r$ and $k'_t$, respectively, we can
expand $\xi_{\bk'} = v_F k'_r + \frac{{k'_t}^2}{2m}$ and
$e_{\bk' - \bk_F}^1 = v_F k'_r + \frac{{k'_t}^2}{4m}$ as in
the case with only one hot spot pair.
If the second $2k_F$-line crosses the first one under an angle
$\phi$, one simply has to rotate the momentum variables to
obtain
\begin{equation}
 e_{\bk' - \bk_F}^2 = v_F (k'_r \cos\phi - k'_t \sin\phi) +
 \frac{(k'_t \cos\phi + k'_r \sin\phi)^2}{4m}  .
\end{equation}
Since the integral is again dominated by contributions with 
$|k'_r| \ll |k'_t|$, we can simplify this to
$e_{\bk' - \bk_F}^2 = - v_F k'_t \sin\phi$.
Changing integration variables to $k' = - e_{\bk'-\bk_F}^1/v_F$
and $\om' = \xi_{\bk'}$, as previously, yields
\begin{eqnarray} \label{Sg_two_1}
 \Im\Sg(\bk_F^1,\om) &=& \frac{M v_F}{N\pi} \int_0^{\om} d\om'
 \int_{-\om'/v_F}^{\infty} \frac{dk'}{\sqrt{\om' + v_F k'}}
 \nonumber \\
 &\times& \sum_{k'_t = \pm 2\sqrt{m} \sqrt{\om' + v_F k'}} 
 \nonumber \\
 &\times& \Im \Big[
 \big( \sqrt{\om' - \om - v_F k'} + \om \lra \om' \big) 
 + \bar b v_F k' 
 \nonumber \\ 
 &+& \big( \sqrt{\om' - \om - v_F k'_t \sin\phi} + \om \lra \om'
 \big) 
 \nonumber \\
 &+& \bar b v_F k'_t \sin\phi \Big]^{-1}
\end{eqnarray}
for $\om > 0$.
Contributions from $k'_t > 0$ dominate over those from $k'_t < 0$ 
for small $\om$.
The first and the last term in the denominator is of order 
$\sqrt{\om}$, while the second term is of order $\om$, and the
third one is of order $\om^{3/4}$ for $k'_t > 0$.
Keeping only the leading terms, and substituting $k' = \om\tk/v_F$,
$\om' = \om\tom$, yields
\begin{eqnarray} \label{Sg_two_2}
 \Im\Sg(\bk_F^1,\om) &=& \frac{M\om}{N\pi} \int_0^1 d\tom
 \int_{-\tom}^{\infty} \frac{d\tk}{\sqrt{\tom + \tk}}
 \nonumber \\
 &\times& \Im \Big[
 \sqrt{\tom - 1 - \tk} + \sqrt{1 - \tom - \tk}
 \nonumber \\
 &+& 2 \bar b v_F \sin\phi \sqrt{m} \sqrt{\tom + \tk} \, \Big]^{-1} .
 \hskip 5mm
\end{eqnarray}
Using $\tilde\kappa = \tom + \tk$ and $\tom$ as integration variables,
implementing also the case $\om < 0$, and restoring the imaginary
infinitesimals $i0^+$, we obtain the final result
\begin{eqnarray} \label{Sg_two_3}
 \Im\Sg(\bk_F^1,\om) &=& \frac{M\om}{N\pi} \int_0^1 d\tom
 \int_{0}^{\infty} \frac{d\tilde\kappa}{\sqrt{\tilde\kappa}}
 \nonumber \\
 &\times& \Im \Big[
 \sqrt{(2\tom - 1)\sgn(\om) + i0^+ - \tilde\kappa}
 \nonumber \\
 &+&  \sqrt{\sgn(\om) - i0^+ - \tilde\kappa} 
 + 2 \tilde b \sqrt{\tilde\kappa} \, \Big]^{-1} , 
 \hskip 5mm
\end{eqnarray}
with the dimensionless constant
\begin{equation}
 \tilde b = \left(v_F \sqrt{m} \sin\phi \right) \bar b = 
 \left( 4\pi N^{-1} v_F^2 \sin\phi \right) b \; . 
\end{equation}
The imaginary part of the self-energy is thus linear in $\om$
at small $\om$, with a prefactor depending only on $\tilde b$.
Note that $\tilde b$ does not depend on the curvature of the
Fermi surface at the hot spot.
A striking feature is that the prefactor for $\om > 0$ differs
from the one for $\om < 0$. For a convex Fermi surface, as 
assumed above, the prefactor for negative energies (holes) is 
larger than for positive energies (particles), and vice versa 
for a concave Fermi surface.
For $\tilde b \gg 1$, Eq.~\ref{Sg_two_3} can be simplified to
$ \Im\Sg(\bk_F^1,\om) = - M N^{-1} C_{\pm} |\om|$ with 
$C_{+} = \big( \frac{1}{4} - \frac{1}{2\pi} \big) {\tilde b}^{-1}$ and
$C_{-} = \big( \frac{1}{4} + \frac{1}{2\pi} \big) {\tilde b}^{-1}$, for
$\om > 0$ and $\om < 0$, respectively. 
These asymptotic expressions provide a good approximation for 
$\tilde b \geq 10$.

Following the above derivation, one can see that including a 
tangential momentum dependence with a prefactor $c$ as in 
Eq.~(\ref{pi0}) merely amounts to an additional contribution
proportional to $k'_t$ in the denominator of Eq.~(\ref{Sg_two_1}),
and subleading terms, so that $\tilde b$ in 
Eq.~(\ref{Sg_two_3}) is replaced by $\tilde b - \tilde c$ with 
$\tilde c = 4\pi N^{-1} v_F (1 - \cos\phi) c$.

Let us pick the $2k_F$ spin-density wave instability obtained 
from Hartree-Fock studies of the two-dimensional Hubbard model 
\cite{schulz90,igoshev10} as an example.
In this case, the constants $\tilde b$ and $\tilde c$ 
can be determined by computing $\Pi_0(\bq,0)$ for $\bq$ near
$\bQ = (\pi,Q)$ and $v_F$ explicitly from the tight-binding 
band structure.
Typical values for $\tilde b$ are between $10$ and $20$, while
$\tilde c$ is considerably smaller. The prefactor to the linear
energy dependence of $\Im\Sg(\bk_F^1,\om)$ is thus about $0.05$
for $\om < 0$ and about $0.01$ for $\om > 0$.

For momenta away from the hot spots, the self-energy obeys
Fermi liquid behavior in the low energy limit. Close to the
hot spots, there is a crossover between the non-Fermi 
liquid power-laws derived above at intermediate energies,
and Fermi liquid behavior at very low energies.
For momenta on the Fermi surface at a small distance $k_t$ 
from the next hot spot, the crossover scale $\om^*$ is 
proportional to $k_t^3$ for the case of a single hot spot pair,
and proportional to $k_t^2$ for two hot spot pairs.

We have computed the self-energy only to one-loop order.
This suffices to detect the breakdown of Fermi liquid theory,
but higher orders may modify the power-laws. For the 
intensively studied cases of commensurate antiferromagnetic 
and nematic quantum criticality, corrections to the one-loop 
power-laws have been found at two-loop \cite{metlitski10_af} 
and three-loop order \cite{metlitski10_nem}, respectively.
Higher order terms may also destroy the QCP at low energy 
scales, and preempt it by a first order transition 
\cite{altshuler95}, or trigger another instability such as 
pairing.

%%% Summary %%%%%%%%%%%%%%%%%%%%%%%%%%%%%%%%%%%%%%%%%%%%%%%%%%%%

In summary, we have analyzed the fate of single-particle
excitations at the onset (QCP) of incommensurate $2k_F$ charge 
or spin density wave order in a two-dimensional metal. 
There are two qualitatively distinct important scenarios, 
namely those with a single pair of hot spots at high lattice 
symmetry positions, and cases with two hot spot pairs.
The dynamical fluctuation propagator exhibits peculiar square 
root singularities in both cases.
The energy dependence of the single-particle decay rate at the 
hot spots as obtained from the one-loop self-energy obeys
non-Fermi liquid power laws, with an exponent 2/3 in the case 
of a single hot spot pair, and exponent one for two hot spot
pairs. The prefactors of the linear behavior obtained in the 
latter case exhibit a pronounced particle-hole asymmetry.

In future work one should analyze the role of higher order 
contributions in a suitable renormalization group framework.
It will also be very interesting to extend the present analysis 
to the quantum critical regime at finite temperature, study 
transport properties, and relate the results to correlated
electron compounds with $2k_F$ instabilities.

%%%%%%%%%%%%%%%%%%%%%%%%%%%%%%%%%%%%%%%%%%%%%%%%%%%%%%%%%%%%%%%%

\begin{acknowledgments}
We are grateful to F.~Benitez, A.~Eberlein, and H.~Yamase for 
valuable discussions.
\end{acknowledgments}

%%%%%%%%%%%%%%%%%%%%%%%%%%%%%%%%%%%%%%%%%%%%%%%%%%%%%%%%%%%%%%%%


\begin{thebibliography}{99}
\bibitem{fn1} Hot spots are points on the Fermi surface that
 are connected by the ordering wave vector $\bQ$.
\end{thebibliography}
\end{document}